\documentclass[conference]{IEEEtran}

\usepackage{xcolor,soul,framed} 

\colorlet{shadecolor}{yellow}
\usepackage[pdftex]{graphicx}
\graphicspath{{../pdf/}{../jpeg/}}
\DeclareGraphicsExtensions{.pdf,.jpeg,.png}

\usepackage[cmex10]{amsmath}
\usepackage{array}
\usepackage{mdwmath}
\usepackage{mdwtab}
\usepackage{eqparbox}
\usepackage{url}
\usepackage{listings}
\usepackage{graphicx}
\usepackage{program}
\usepackage{ifthen}
\usepackage{epsfig}
\usepackage{array}
\usepackage{color}
\usepackage{hyperref}
\usepackage{url}
\usepackage{xspace}
\usepackage{wasysym}
\usepackage{booktabs}
\usepackage{algorithmic}
\usepackage{textcomp}
\usepackage{comment}

\usepackage{subcaption}
\definecolor{blue}{rgb}{0,0,1.0}
\definecolor{darkgreen}{rgb}{0,0.44,0}
\definecolor{green}{rgb}{0,0.44,0}
\definecolor{darkred}{rgb}{0.44,0,0}
\definecolor{darkblue}{rgb}{0,0,0.64}
\definecolor{mygray}{rgb}{0.9,0.9,0.9}
\definecolor{mymauve}{rgb}{0.58,0,0.82}
\definecolor{myred}{rgb}{0.72,0.18,0.0} 
\definecolor{mygreen}{rgb}    {0.0,0.72,0.0} 
\definecolor{myblue}{rgb} {0.18,0.0,0.72} 
\definecolor{mycreme}{rgb}        {1.0,0.8,0.2} 
\definecolor{mygray}{rgb}{0.95,0.95,0.95}
\definecolor{darkgray}{rgb}{0.55,0.55,0.55}

\newcommand{\gemm}{{\sc gemm}\xspace}

\newcommand{\imr}{{\sc im2row}\xspace}
\newcommand{\exo}{{\sc Exo}\xspace}

\begin{document}
\bstctlcite{IEEEexample:BSTcontrol}
    \title{Tackling the Matrix Multiplication Micro-kernel Generation with \exo}
 
 \author{\IEEEauthorblockN{Adri\'an Castell\'o }
\IEEEauthorblockA{
\textit{Universitat Polit\`ecnica de Val\`encia}\\
{adcastel@disca.upv.es}}
\and
\IEEEauthorblockN{Julian Bellavita}
\IEEEauthorblockA{\textit{Cornell University} \\
jb2695@cornell.edu}
\and
\IEEEauthorblockN{Grace Dinh}
\IEEEauthorblockA{\textit{UC Berkeley} \\
dinh@berkeley.edu}
\and
\IEEEauthorblockN{Yuka Ikarashi}
\IEEEauthorblockA{\textit{MIT CSAIL} \\
yuka@csail.mit.edu}
\and
\IEEEauthorblockN{H\'ector Mart\'inez}
\IEEEauthorblockA{\textit{Universidad de C\'ordoba} \\
el2mapeh@uco.es}
  }


\maketitle

\begin{abstract}
The optimization of the matrix multiplication (or \gemm) has been a need during the last decades. This operation is considered the flagship of current linear algebra libraries such as BLIS, OpenBLAS, or Intel OneAPI because of its widespread use in a large variety of scientific applications. The \gemm is usually implemented following the GotoBLAS philosophy, which tiles the \gemm operands and uses a series of nested loops for performance improvement. These approaches extract the maximum computational power of the architectures through small pieces of hardware-oriented, high-performance code called micro-kernel. However, this approach forces developers to generate, with a non-negligible effort, a dedicated micro-kernel for each new hardware.

In this work, we present a step-by-step procedure for generating micro-kernels with the \exo compiler that perform close to (or even better than) manually developed microkernels written with intrinsic functions or assembly language. Our solution also improves the portability of the generated code, since a hardware target is fully specified by a concise library-based description of its instructions. 
\end{abstract}


\section{Introduction}
\label{sec:introduction}
%
%

%
%
%


Over the past few decades, there has been a relentless effort to develop  high-performance implementations
of linear algebra (LA) libraries. These solutions have been designed to target a wide variety of architectures, such as vector processors,
multi-core processors,
data-parallel graphics processing units (GPUs), and, more recently, 
domain-specific architectures and accelerators.
This not-insignificant effort usually comes from major hardware vendors, with some relevant products being Intel OneAPI, 
AMD AOCL, IBM ESSL, ARMPL, and NVIDIA cuBLAS, as well as from the academic side, with software packages such
as GotoBLAS2~\cite{Goto:2008:AHP}, 
OpenBLAS~\cite{OpenBLAS:ICPADS}, 
BLIS~\cite{BLIS1} and
ATLAS~\cite{ATLAS}. 

The general
matrix multiplication (\gemm) is the flagship computational kernel on which these LA libraries 
are built upon. 
{\color{black} The overall performance of \gemm comes from the macro-kernel which comprises a series of general code optimizations such as tiling, loop reordering, and data packing, and the micro-kernel which is responsible for a significant part of the performance because is the link between the \gemm algorithm and the underlying hardware.}
In addition, \gemm is also a key operation for deep learning (DL) applications that use convolutional deep neural networks (DNNs)
for signal processing and computer vision~\cite{8114708,DBLP:journals/corr/abs-1802-09941}. 
Unfortunately, these LA libraries have some limitations:
\begin{enumerate}
\item The optimized routines are hardware-specific. This is the case for Intel, IBM, or ARM products. To a lesser, it also applies to GotoBLAS2, OpenBLAS, and BLIS, which use a collection of hardware-oriented micro-kernels \cite{BLIS2}.
\item Developing highly optimized, hand-written, micro-kernels for \gemm requires deep knowledge of high-performance computing and computer architecture.
\item A new architecture implies a new set of test-and-debug development of  micro-kernels to extract the maximum computational power from the hardware.
\item The code of these micro-kernels usually uses productivity-enhancing 
      macros, templates, and 
      high-level programming techniques. Therefore, maintaining the libraries thus mostly
      lies in the hands of the original developers.
 
\item The software {misses some relevant 
      cases} such as for example, support for half (i.e., 16-bit) floating point precision or integer arithmetic.

{\color{black}
\item The implementation of a unique micro-kernel for all \gemm scenarios may incur in a performance loss for non-squarish \gemm as they appear in DL. 
}
 
\end{enumerate}

In this paper, we address these limitations by demonstrating that it is possible to automatically generate a set of micro-kernels for \gemm using \exo~\cite{yukaexo}.  
This alternative solution has the following advantages:
\begin{enumerate}

 \item At a high level, the library micro-kernel is ``replaced'' by a collection of \exo generated C code. Each micro-kernel will handle a different edge case.
 \item Using the appropriate backend, 
       the generation/optimization can be 
       easily specialized for different data types,  which 
       further enhances the portability and maintainability of the solution.
 \item By matching  the size of the micro-kernel  
       to the problem, it is possible
       to outperform the high-performance realizations of \gemm of existing, widely-accepted solutions.
 \item The optimization process for each problem is greatly reduced, boiling down to  evaluating  a number of generated micro-kernels.
 \item The micro-kernel generator for ARM Neon is publicly available at \url{https://github.com/adcastel/EXO_ukr_generator}.

\end{enumerate}
{\color{black} Moreover, this work highlights the usability of \exo by showing that the performance achieved by auto-generated C code is as good as hand-written solutions by generating different codes for matching the problem sizes.}

In addition, the work in this paper has contributed to the \exo code with the support for two Neon intrinsic instructions and the support for 16-bit floating point data types for ARM\footnote{FP16 support is not in the \exo repository at the time of writing this paper but is available at \url{https://github.com/adcastel/exo}}.

The rest of the paper is organized as follows. Section~\ref{sec:back} resumes the BLIS algorithm for \gemm and introduces the \exo domain-specific programming language; Section~\ref{sec:related} 
visits some existing work; Section~\ref{sec:generator} presents a step-by-step process of how to generate optimized micro-kernel ARM codes using \exo; Section~\ref{sec:experiments} evaluates and compares the generated micro-kernels with other micro-kernels in three different scenarios; and Section~\ref{sec:conclusions} summarizes the work done and its contributions.

\section{Background}
\label{sec:back}
\subsection{BLIS algorithm for \gemm}
\begin{figure}[tb!]
\centering
\includegraphics[width=0.75\columnwidth]{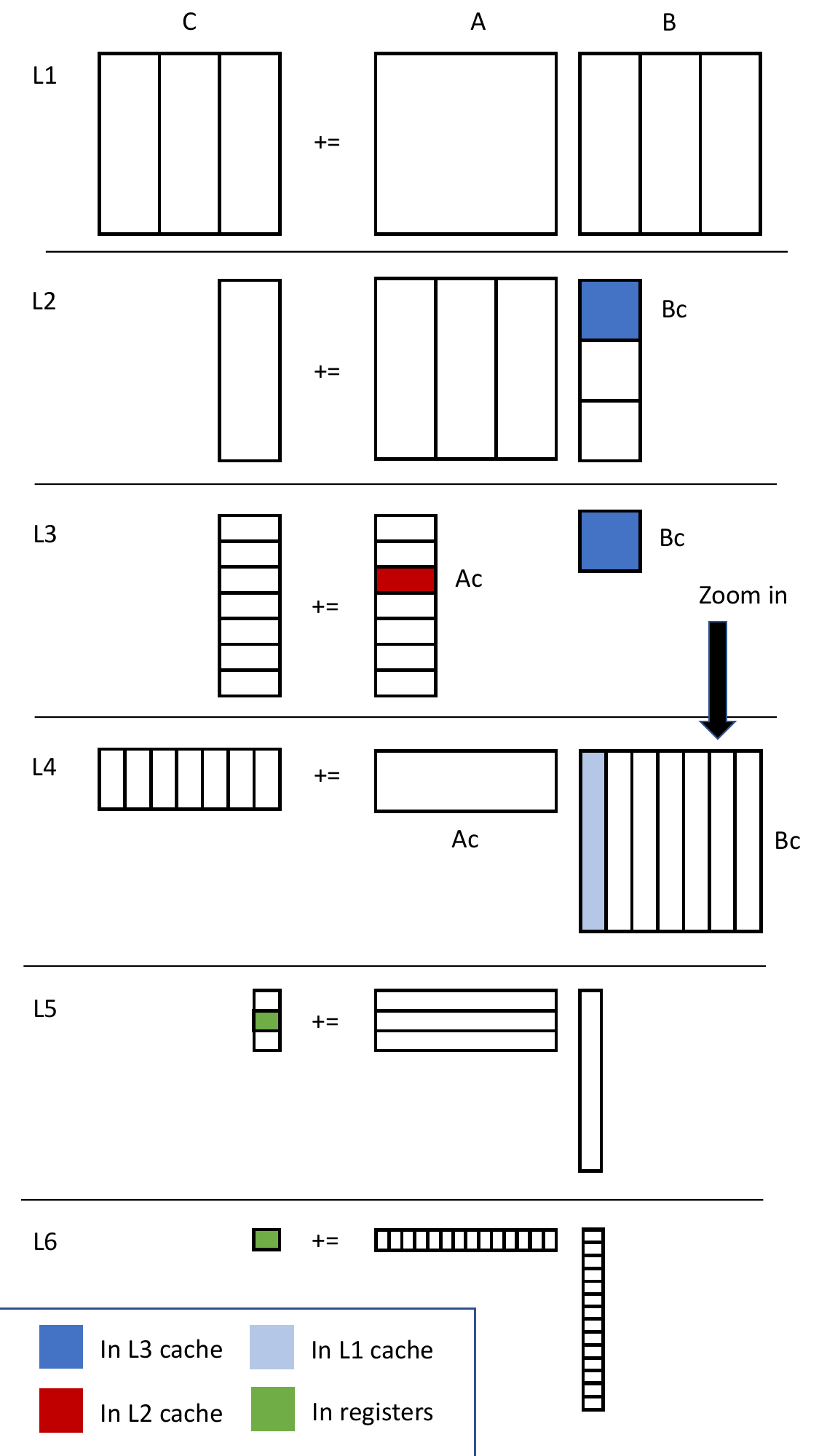}
\caption{Overview of the BLIS base \gemm algorithm.}
\label{fig:blis_family_Creg}
\end{figure}

Consider the \gemm $C = C + AB$, where the operands are matrices with the following dimensions: $m \times k$, $k \times n$, and $m \times n$ for $A$, $B$, and $C$, respectively.
The BLIS framework (as well as other LA libraries) follows GotoBLAS~\cite{Goto:2008:AHP} approach to encoding this operation as three nested loops around two \textit{packing routines}. {\color{black} This structure is known as macro-kernel and its code is usually shared across different hardware architectures.} 
In BLIS, the macro-kernel is decomposed into two additional loops around a \textit{micro-kernel}, with the latter consisting of a single loop that performs one {outer product} per iteration.
The Figures~\ref{lst:bliscode} and~\ref{fig:blis_family_Creg} 
show the \textit{BLIS baseline algorithm} 
for \gemm,  which includes the six loops, the two packing routines,
and the micro-kernel. 

\begin{figure}[tbh!]
\centering
\begin{minipage}[t]{0.9\columnwidth}
\begin{lstlisting}[language=C,alsoletter={.},deletekeywords={.sum,in},morekeywords={@proc,DRAM, f32, divide_loop}]
for (jc=0; jc<n; jc+=nc)  // Loop L1
 for (pc=0; pc<k; pc+=kc) {    // L2
  // Pack B
  Bc := B(pc:pc+kc-1,jc:jc+nc-1);
  for (ic=0; ic<m; ic+=mc) {   // L3
   // Pack A
   Ac := A(ic:ic+mc-1,pc:pc+kc-1);
   for (jr=0; jr<nc; jr+=nr)   // L4
     for (ir=0; ir<mc; ir+=mr) // L5
      // Micro-kernel          // L6
      C(ic+ir:ic+ir+mr-1,      
        jc+jr:jc+jr+nr-1)
        += Ac(ir:ir+mr-1,0:kc-1)
        *  Bc(0:kc-1,jr:jr+nr-1);
  }}
\end{lstlisting}
\end{minipage}
\caption{Pseudo-code of the BLIS base \gemm algorithm.}
\label{lst:bliscode} 
\end{figure}

The three outermost loops of the algorithm travel the 
$n$--,
$k$--, and
$m$--dimensions of the problem, partitioning the matrix operands to fit  the processor's
cache hierarchy. The cache configuration parameters
$n_c$, 
$k_c$, and  
$m_c$, adapted to the target processor architecture~\cite{BLIS4}, 
favor that 
$B_c$ stays in the L3 cache and
$A_c$ in the L2 cache during the execution of the micro-kernel,
while $C$ is streamed from the main memory into the processor registers as shown in Figure~\ref{fig:blis_family_Creg}

{\color{black}In addition, the macro-kerenl packing routines ensure that the data in $A_c, B_c$ is accessed with unit stride from the micro-kernel.
}

The micro-kernels are pieces of hardware-oriented code that are usually
encoded using assembly language. 
Inside the micro-kernel, a block of $m_r \times n_r$ elements of $C$ is updated inside the $k_c$ loop. The operands for the update are portions of $A_c$ and $B_c$ buffers, more specifically $A_r$ and $B_r$, whose dimensions are $m_r\times k_c$ and $k_c \times n_r$, respectively. The values of $m_r$ and $n_r$ are commonly used to name the micro-kernel. 
BLIS provides specialized micro-kernels for many processors from AMD, Intel, ARM, and IBM. However,  developing a new hardware-oriented micro-kernel for each new architecture is an expensive development effort, so BLIS provides one micro-kernel per architecture using a monolithic approach.

\subsection{Compilation and Tuning Frameworks}
\label{sec:exo}

The expense of developing optimized micro-kernels for diverse architectures largely stems from the need to perform architecture-specific optimizations to obtain maximum performance, 
 and to generate platform-specific instructions such as hardware intrinsics. To address this, \emph{automatic code generation} approaches have been applied to achieve performance portability across existing and new 
architectures, including general-purpose processors or specific-purpose accelerators~\cite{MoreauVTA2018, GKA+21-Gemmini}, with a minimum of programmer intervention~\cite{DBLP:journals/corr/abs-2002-03794}.

Specifically, \emph{user-schedulable} languages and compiler frameworks, such as Halide~\cite{Halide}, MLIR~\cite{mlir}, TVM~\cite{TVM_1} or \exo~\cite{yukaexo}, propose a clear separation of concerns between  
the definition of an operation (e.g. a matrix multiplication) and the \emph{schedule}, or set of optimizations applied to the operation, allowing common optimizations (e.g loop tiling for higher levels of the memory hierarchy) to be abstracted away from architecture-specific optimizations such as vectorization. 

Compiler frameworks must also be given a \emph{hardware specification} describing the hardware's instruction set (including vector intrinsics). Traditional compilers such as Halide, TVM, and LLVM integrate hardware specifications tightly into the compiler, requiring manual specification of code generation passes to support new hardware. {\color{black} Therefore, sometimes, the generation tools are bound to certain compilers so the user cannot deal with several code decisions. As an example, TVM and LLVM are bound so the user does not have to compile the code but, at the same time, there are some aspects that the user is not aware of.}

\exo~\cite{yukaexo}, on the other hand, externalizes the definition of hardware intrinsics by taking as input user-defined libraries. For instance, Fig. \ref{lst:hwconfig} provides definitions for the ARM Neon intrinsic functions \texttt{vstlq\_f32} and \texttt{vfmaq\_laneq\_f32} in semantically equivalent Python; these definitions are used by the compiler to automatically generate calls. An entire hardware specification consists of similar definitions for each hardware intrinsic, as well as definitions for levels of the memory hierarchy (in this case, DRAM and Neon registers) and datatypes (in this case, \texttt{f32} floats). {\color{black}Concretely, these definitions will ensure that the user methods do not change the behavior of the original code by checking the intrinsic replacement with the expected pattern. Without this security definition, the user could change any loop by any intrinsic instruction which may evoke a different code.  Another key aspect from \exo is its independence from any compiler\slash optimizer tool. Specifically, \exo ``just'' generate a C code with intrinsic instructions that need to be compiled and, therefore, the user can try different combinations of hardware\slash compiler to obtain the maximum performance of the generated code.}

\begin{figure}[tbh!]
\centering
\begin{minipage}[t]{0.9\columnwidth}
\begin{lstlisting}[language=python,alsoletter={.},deletekeywords={.sum,in},morekeywords={@proc,DRAM, f32, assert, @instr, neon_vst_4xf32, neon_vfmla_4xf32_4xf32, Neon, index}]
@instr("vst1q_f32(&{dst_data}, {src_data});")
def neon_vst_4xf32(dst: [f32][4] @ DRAM, 
                   src: [f32][4] @ Neon):
  assert stride(src, 0) == 1
  assert stride(dst, 0) == 1

  for i in seq(0, 4):
    dst[i] = src[i]

@instr("{dst_data} = vfmaq_laneq_f32(
        {dst_data}, {lhs_data}, {rhs_data}, {l});")
def neon_vfmla_4xf32_4xf32(dst: [f32][4] @ Neon, 
                           lhs: [f32][4] @ Neon, 
                           rhs: [f32][4] @ Neon, 
                           l: index):
  assert stride(dst, 0) == 1
  assert stride(lhs, 0) == 1
  assert stride(rhs, 0) == 1
  assert l >= 0
  assert l < 4
  
  for i in seq(0, 4):
    dst[i] += lhs[i] * rhs[l]
\end{lstlisting}
\end{minipage}
\caption{Hardware specification (specified as a library) for vector intrnsic instructions in \exo}
\label{lst:hwconfig} 
\end{figure}

A serious drawback of some of these frameworks is the reliance on existing libraries. For example, the code generated by TVM uses a set of TVM objects that may force rewriting a large part (or all) of the software stack, in addition to incurring runtime overheads to convert datatypes to those supported by these libraries. 
However, tools such as \exo generate plain C code that can be used within all the existing performance-oriented libraries. 

\subsection{Autotuning and Optimization}
\label{sec:related}
By treating schedules as inputs user-schedulable languages facilitate the automatic exploration of scheduling spaces through auto-tuning techniques~\cite{MAS+16-halideautoschedule,ZJS+20Ansor}. For example,
AutoTVM~\cite{Chen18_Learning}, as part of TVM, performs a full exploration of a search space defined by a user-specified template. While such search methods have shown strong performance in practice, they must contend with a computationally expensive search space of possible tuning parameters that grows exponentially with the dimension of the design space as well as difficulties in generalizability (both to different problem sizes and to different hardware targets) and explainability to developers.


Recently, work in \cite{BLIS4} as part of the BLIS project, has shown that 
the use of analytical models for optimal 
configuration parameters selection is an effective way to achieve high performance without the need for auto-tuning. 
Replacing the auto-tuning scheme with model-based solutions
has also been successfully explored in~\cite{OpenBLAS:ICPADS,Yotov:2005,OIT+21-IOOpt,HKD+21}.

This work, focused on micro-kernel generation, differs from~\cite{GEMM_MLIR}, which uses MLIR to describe early experiences with the entire \gemm algorithm, from \cite{Zhang2009}, which proposes advanced auto-tuning schemes for the primitive, and from~\cite{alaejos2022micro} which uses a Python script and C macros to build micro-kernels. Specifically, we use \exo to extend and further analyze the manual generation of \gemm micro-kernels and integrate the resulting code into a BLIS-like \gemm algorithm.

\section{Code Generation}
\label{sec:generator}
In this section, we  explain in a step-by-step manner how to build an HPC micro-kernel code with \exo for ARMv8 architecture from scratch. The dimensions of the micro-kernel will be $8 \times 12$, as the one present in the BLIS library for this specific architecture. For each building phase, we  first introduce the employed \exo's instructions used and explain the resulting intermediate code. We  also show how to extend the code generation to handle other features of the micro-kernels. The code for the step-by-step generation is available at \url{https://github.com/adcastel/EXO_ukr_generator}.

\subsection{Micro-kernel Generation}
Figure~\ref{lst:exo_ukernel} shows what the initial code looks like for a column-major micro-kernel based on the outer product (as in BLIS). To meet the BLIS \gemm algorithm features, we have modified the initial micro-kernel code as follows: 1) Since the C language is a row-major allocator, we  transpose the $C$ matrix dimensions; 2) BLIS uses  data packings for the $A$ and $B$ operands  of the micro-kernel, which guarantees unit stride access to the data. Therefore, to have this in the $A_c$ operand, we also swap the $A_c$ dimensions. The $B_c$ operand is already accessed with a unit stride so no changes are needed; 3) We  rearrange the internal loops in a $k,j,i$ order to match the desired structure. Note that $A_c$ and $B_c$ operands are allocated in a 1-dimensional structure in the algorithm that calls the micro-kernel, so although we transpose the data buffers $C$ and $A_c$ there is no risk of the access pattern.

To start with the explanation of the code in Figure~\ref{lst:exo_ukernel}, the first line (\texttt{@proc}) tells the \exo compiler that the next function is schedulable, and will therefore produce a C-language code. The arguments of the function also show some aspects of the \exo language. First, the argument name  is followed by ``$:$'' and the type, which can be \texttt{size, scalar, vector}, or \texttt{matrix}. For data that requires memory allocation, we should specify its placement. In this example, we map the $A_c$, $B_c$, and $C$ buffers to RAM using the \texttt{@ DRAM} notation. 
This version covers all combinations of alpha and beta values. The buffers $C_b$ and $B_a$ located at lines 8 and 9 are used for the computation of $C * beta$ and $B_c * alpha$. Specifically, lines 12--14 compute the $C_b$ results and lines 17--19 compute the $B_a$ results.

Lines 22--25 perform the micro-kernel result by executing $C_b = C_b + A_c \times B_a$. Finally, lines 28--30 will return the result of the computation to the $C$ matrix.

\begin{figure}[tbh!]
\centering
\begin{minipage}[t]{0.9\columnwidth}
\begin{lstlisting}[language=python,alsoletter={.},deletekeywords={.sum,in},morekeywords={@proc,DRAM, f32}]
@proc
def ukernel_ref( MR: size, NR: size, KC: size, 
  alpha: f32[1], Ac: f32[KC, MR] @ DRAM, 
  Bc: f32[KC, NR] @ DRAM, beta: f32[1], 
  C: f32[NR, MR] @ DRAM,
  ):
  # Tmp buffers for C * beta and B * alpha
  Cb: f32[NR,MR] @ DRAM
  Ba: f32[KC,NR] @ DRAM

  # Cb = C * beta
  for cj in seq(0, NR):
    for ci in seq(0, MR):
      Cb[cj,ci] = C[cj,ci] * beta[0]
       
  # Ba = Bc * alpha
  for bk in seq(0, KC):
    for bj in seq(0, NR):
      Ba[bk,bj] = Bc[bk,bj] * alpha[0]

  # C += Ac * Bc
  for k in seq(0, KC):
    for j in seq(0, NR):
      for i in seq(0, MR):
        Cb[j, i] += Ac[k,i] * Ba[k,j]

  # C = Cb
  for cj in seq(0, NR):
    for ci in seq(0, MR):
      C[cj,ci] = Cb[cj,ci]

\end{lstlisting}
\end{minipage}
\caption{\exo code for \gemm micro-kernel.}
\label{lst:exo_ukernel} 
\end{figure}

For simplicity, from this point on, we will optimize a specific version of the micro-kernel. Specifically, we will apply the step-by-step transformation to the code shown in Figure~\ref{lst:exo_ukernel_simplified} that corresponds to the micro-kernel when $alpha$ and $beta$ values both equal 1. 

Optimization of the initial code will involve more scheduling functions for the $C_b$ and $B_a$ loops (lines 11--14, 17--19, and 27--30), equivalent to those 
 shown from this point beyond.

\begin{figure}[tbh!]
\centering
\begin{minipage}[t]{0.9\columnwidth}
\begin{lstlisting}[language=python,alsoletter={.},deletekeywords={.sum,in},morekeywords={@proc,DRAM, f32}]
@proc
def ukernel_ref( MR: size, NR: size, KC: size, 
  alpha: f32[1], Ac: f32[KC, MR] @ DRAM,
  Bc: f32[KC, NR] @ DRAM, beta: f32[1],
  C: f32[NR, MR] @ DRAM,
  ):

  # C += Ac * Bc
  for k in seq(0, KC):
    for j in seq(0, NR):
      for i in seq(0, MR):
        C[j, i] += Ac[k,i] * Bc[k,j]


\end{lstlisting}
\end{minipage}
\caption{Simplified \exo code for \gemm micro-kernel.}
\label{lst:exo_ukernel_simplified} 
\end{figure}

\paragraph{\textbf{Basic Micro-kernel}} First, we  generate the function to schedule and we  specialize the generated code by specifying that we want to use the values of 8 and 12 for the $M_R$ and $N_R$ arguments, respectively. {\color{black} This conversion is done via the \texttt{partial\_eval} function that replace the values $M_R$ and $N_R$}. Figure~\ref{lst:v1} shows the user code and the generated representation. Lines 3 and 4 get the initial version of the micro-kernel and set the variables ${M_R}$ and $N_R$ variables, respectively. The generated code has changed the variables by their values (e.g. the iterations of the second loop now go from 0 to 12 instead of $N_R$).

\begin{figure}[tbh!]
\centering
\begin{minipage}[t]{0.9\columnwidth}
\begin{lstlisting}[language=python,alsoletter={.},deletekeywords={.sum,in,p.},morekeywords={@proc,DRAM, f32, rename, p.partial_eval}]
# USER CODE
MR, NR = 8, 12
p = rename(ukernel_ref, "uk{}x{}".format(MR,NR))
p = p.partial_eval(MR,NR)

# RESULTING EXO GENERATED CODE
def uk_8x12( KC: size, alpha: f32[1] @ DRAM, 
  Ac: f32[KC, 8] @ DRAM, Bc: f32[KC, 12] @ DRAM, 
  beta: f32[1] @ DRAM, C: f32[12, 8] @ DRAM
  ):
    
  # C += Ac * Bc
  for k in seq(0, KC):
    for j in seq(0, 12):
      for i in seq(0, 8):
        C[j, i] += Ac[k, i] * Bc[k, j]
        
\end{lstlisting}
\end{minipage}
\caption{\exo code for \gemm micro-kernel v1.}
\label{lst:v1} 
\end{figure}

\paragraph{\textbf{Loop Structure}} In lines 2 and 3 of Figure~\ref{lst:v2} we split both the \texttt{i} and \texttt{j} loops to match the vector length of the architecture. As the ARM-based NVIDIA Carmel processor uses 128-bit vector registers and, with float32 data type, the vector length is 4. In addition, the BLIS-like \gemm algorithm used in this work ensures that the $A_c$ and $B_c$ buffer sizes are multiples of the $M_R$ and $N_R$ values, respectively. This action results in the nested loops \texttt{it, itt, jt} and \texttt{jtt} locate in lines 14--17. In addition, \exo has automatically tiled the access to the $C$, $A_c$, and $B_c$ data to match the new loop structure (lines 18--20). 

\begin{figure}[tbh!]
\centering
\begin{minipage}[t]{0.9\columnwidth}
\begin{lstlisting}[language=python,alsoletter={.},deletekeywords={.sum,in},morekeywords={@proc,DRAM, f32, divide_loop}]
# USER CODE
p = divide_loop(p,'i', 4, ['it','itt'], perfect=True)
p = divide_loop(p,'j', 4, ['jt','jtt'], perfect=True)

# RESULTING EXO GENERATED CODE
def uk_8x12( KC: size, alpha: f32[1] @ DRAM, 
  Ac: f32[KC, 8] @ DRAM, Bc: f32[KC, 12] @ DRAM, 
  beta: f32[1] @ DRAM, C: f32[12, 8] @ DRAM
  ):
    
  # C += Ac * Bc
  for k in seq(0, KC):
    # Loop splits to match the vector length (4)
    for jt in seq(0, 3):
      for jtt in seq(0, 4):
        for it in seq(0, 2):
          for itt in seq(0, 4):
            C[jtt + 4 * jt, itt + 4 * it] += 
              Ac[k, itt + 4 * it] * 
              Bc[k, jtt + 4 * jt]
\end{lstlisting}
\end{minipage}
\caption{\exo code for \gemm micro-kernel v2.}
\label{lst:v2} 
\end{figure}

\paragraph{\textbf{C Matrix}} Figure~\ref{lst:v3} shows one of the most complex steps in the micro-kernel generation, the binding of the $C$ matrix to vectorial registers, which includes: declaration, loading, and storing the results. Specifically, the process is as follows:
\begin{enumerate}
    \item Map the $C$ matrix with a vector register (lines 3 and 4), which is reflected in line 51. {\color{black} The \texttt{stage\_mem} function binds the $C$ operand memory to a vectorial register so \exo can then set the $C$ operand data movement.}
    \item Resize the vectorial register to a 3D structure where each dimension corresponds to  the iterations of each loop. {\color{black}This register resize is done by using the \texttt{expand\_dim} method. }Specifically, the first invocation (line 7) resizes the declaration to the size of the vector length, which in this case is 4; line 8 completes the size of the $M_R$ dimension; and line 9 is bound to the $N_R$ dimension of $C$. The result of these lines can be seen in line 34, where the final allocation appears. 
    \item The \texttt{lift\_alloc} statement moves the declaration of the registers of $C$ to the top of the generated code.
    \item Lines 15--18 move the load and the store of the $C$ matrix out of the computation loop (lines 37--43 and lines 56--62, respectively).
    \item Lines 21 and 22 replace the \texttt{itt} loops of the load and store with intrinsic instructions. Lines 40 and 59 emphasize these replacements. 
    \item Line 25 sets the $C$ register variable to type \texttt{Neon}.
\end{enumerate}

\begin{figure}[tbh!]
\centering
\begin{minipage}[t]{0.9\columnwidth}
\begin{lstlisting}[language=python,alsoletter={.},deletekeywords={.sum,in},morekeywords={@proc,DRAM, f32,Neon,neon_vld_4xf32,neon_vst_4xf32,stage_mem, expand_dim,lift_alloc,autofission, replace, set_memory}]
# USER CODE
#  1) Map C buffer to vectorial register C_reg
Cp = 'C[4 * jt + jtt, 4 * it + itt]'
p = stage_mem(p, 'C[_] += _', Cp, 'C_reg')

#  2) Build a 3D structure of C_reg
p = expand_dim(p, 'C_reg', 4, 'itt', ...)
p = expand_dim(p, 'C_reg', MR//4, 'it', ...)
p = expand_dim(p, 'C_reg', NR, 'jt*4+jtt', ...)

#  3) Move the register declaration to the top
p = lift_alloc(p, 'C_reg', n_lifts=5)

#  4) Extract the C load and store from the k-loop
p = autofission(p, p.find('C_reg[_] = _').after(),
                n_lifts=5)
p = autofission(p, p.find('C[_] = _').before(), 
                n_lifts=5)

#  5) Replace the indicated loops by Neon intrinsics
p = replace(p, 'for itt in _: _', neon_vld_4xf32)
p = replace(p, 'for itt in _: _', neon_vst_4xf32)

#  6) Set the C_reg memory to Neon
p = set_memory(p, 'C_reg', Neon)

# RESULTING EXO GENERATED CODE
def uk_8x12( KC: size, alpha: f32[1] @ DRAM, 
  Ac: f32[KC, 8] @ DRAM, Bc: f32[KC, 12] @ DRAM, 
  beta: f32[1] @ DRAM, C: f32[12, 8] @ DRAM
  ):
  
  # Registers for C  
  C_reg: f32[12, 2, 4] @ Neon
  
  # Load C to registers
  for jt in seq(0, 3):
    for jtt in seq(0, 4):
      for it in seq(0, 2):
        neon_vld_4xf32(
            C_reg[4 * jt + jtt, it, 0:4],
            C[4 * jt + jtt, 4 * it:4 * it + 4]
        )

  # C += Ac * Bc
  for k in seq(0, KC):
    for jt in seq(0, 3):
      for jtt in seq(0, 4):
        for it in seq(0, 2):
          for itt in seq(0, 4):
            C_reg[jt * 4 + jtt, it, itt] += 
              Ac[k, itt + 4 * it] * 
              Bc[k, jtt + 4 * jt]
  
  # Store C from registers
  for jt in seq(0, 3):
    for jtt in seq(0, 4):
      for it in seq(0, 2):
        neon_vst_4xf32(
            C[jtt + 4 * jt, 4 * it:4 * it + 4],
            C_reg[jtt + 4 * jt, it, 0:4]
        )
\end{lstlisting}
\end{minipage}
\caption{\exo code for \gemm micro-kernel v3.}
\label{lst:v3} 
\end{figure}

\paragraph{\textbf{$A_c$ and $B_c$ Operands}} Figure~\ref{lst:v4} lists the actions to generate the loads from $A_c$ and $B_c$  to registers. Note that the code uses the name $X_c$  for simplicity since these actions must be performed for both operands. The procedure for each operand is as follows:
\begin{enumerate}
    \item Map the $X_c$ matrix to a vector register (line 3).
    \item Resize the vector register to a 2D structure where the first dimension is the vector length and the second dimension is the outermost loop (lines 6 and 7). The result of these lines can be seen in lines 32 and 33 where the final allocation appears. 
    \item The \texttt{lift\_alloc} statement moves the declaration of the registers outside the $k$-loop.
    \item Lines 13 and 14 move the loading of the operands to the $k$-loop.
    \item Line 17 replaces the \texttt{xtt} loops with neon vector load instructions.
    \item Line 20 sets the register variable to \texttt{Neon} type.
\end{enumerate}

\begin{figure}[tbh!]
\centering
\begin{minipage}[t]{0.9\columnwidth}
\begin{lstlisting}[language=python,alsoletter={.},deletekeywords={.sum,in},morekeywords={@proc,DRAM, f32,Neon,neon_vld_4xf32,neon_vst_4xf32,stage_mem, expand_dim,lift_alloc,autofission, replace, set_memory,bind_expr}]
# USER CODE
#  1) Map Xc buffer to vectorial register X_reg
p = bind_expr(p, 'Xc[_]','X_reg')

#  2) Build a 2D structure of X_reg
p = expand_dim(p, 'X_reg' , 4, 'xtt', ...)
p = expand_dim(p, 'X_reg', XR//4, 'xt', ...)

#  3) Move the register declaration to the top
p = lift_alloc(p, 'X_reg', n_lifts=5)

#  4) Move the Xc load to the k-loop
p = autofission(p, p.find('X_reg[_] = _').after(),
                n_lifts=4)

#  5) Replace the xtt loop by Neon intrinsics
p = replace(p, 'for xtt in _: _', neon_vld_4xf32)

#  6) Set the X_reg memory to Neon
p = set_memory(p, 'X_reg', Neon)

# RESULTING EXO GENERATED CODE
def uk_8x12( KC: size, alpha: f32[1] @ DRAM, 
  Ac: f32[KC, 8] @ DRAM, Bc: f32[KC, 12] @ DRAM, 
  beta: f32[1] @ DRAM, C: f32[12, 8] @ DRAM
  ):
  
  # Registers for C and load C as in the
  # previous figure.  Omitted for brevity

  # Registers for Ac and Bc
  A_reg: R[2, 4] @ Neon
  B_reg: R[3, 4] @ Neon
  for k in seq(0, KC):
    # Load Ac and Bc to registers
    for it in seq(0, 2):
      neon_vld_4xf32(
        A_reg[it, 0:4], 
        Ac[k, 4 * it:4 + 4 * it])
    for jt in seq(0, 3):
      neon_vld_4xf32(
        B_reg[jt, 0:4], 
        Bc[k, 4 * jt:4 + 4 * jt])
    for jt in seq(0, 3):
      for jtt in seq(0, 4):
        for it in seq(0, 2):
          for itt in seq(0, 4):
            C_reg[jtt + 4 * jt, it, itt] += 
              A_reg[it, itt] * B_reg[jt, jtt]
    
  # Store C from registers as in the
  # previous figure.  Omitted for brevity 
\end{lstlisting}
\end{minipage}
\caption{\exo code for \gemm micro-kernel v4.}
\label{lst:v4} 
\end{figure}

\paragraph{\textbf{\gemm Operation}} Figure~\ref{lst:v5} illustrates this step. We  reorder the \texttt{jtt} and \texttt{it} loops of the calculation (line 2) so that the access to the $B$ register values is sequential. Line 3 replaces the innermost loop for the \texttt{fmla} statement  as  shown in lines 24--26. 

\begin{figure}[tbh!]
\centering
\begin{minipage}[t]{0.9\columnwidth}
\begin{lstlisting}[language=python,alsoletter={.},deletekeywords={.sum,in},morekeywords={@proc,DRAM, f32,Neon,neon_vld_4xf32,neon_vst_4xf32,stage_mem, expand_dim,lift_alloc,autofission, replace, set_memory,bind_expr,reorder_loops,neon_vfmla_4xf32_4xf32}]
# USER CODE
p = reorder_loops(p,'jtt it')
p = replace(p, 'for itt in _: _', 
                neon_vfmla_4xf32_4xf32)
      
# RESULTING EXO GENERATED CODE
def uk_8x12( KC: size, alpha: f32[1] @ DRAM, 
  Ac: f32[KC, 8] @ DRAM, Bc: f32[KC, 12] @ DRAM, 
  beta: f32[1] @ DRAM, C: f32[12, 8] @ DRAM
  ):
  
  # Registers for C, Ac and Bc and loads as in 
  # previous figures.  Omitted for brevity

  # C += Ac * Bc 
  for k in seq(0, KC):
    # Load Ac and Bc to registers as in 
    # previous figures. Omitted for brevity
    
    # Computation with registers
    for jt in seq(0, 3):
      for it in seq(0, 2):
        for jtt in seq(0, 4):
          neon_vfmla_4xf32_4xf32(
            C_reg[jtt + 4 * jt, it, 0:4],
            A_reg[it, 0:4], B_reg[0:4, jt], jtt)
  
  # Store C from registers as in 
  # previous figures.  Omitted for brevity 
\end{lstlisting}
\end{minipage}
\caption{\exo code for \gemm micro-kernel v5.}
\label{lst:v5} 
\end{figure}

\paragraph{\textbf{Loop Unrolling}} Although this is a technique that some compilers use by default, it is also possible to do it manually in \exo. Figure~\ref{lst:v6} shows an example of unrolling for the loops that load $A_c$ and $B_c$ operands into registers. We show it with the \texttt{unroll\_loop} statements of lines 2 and 3, resulting in lines 21--25.

\begin{figure}[tbh!]
\centering
\begin{minipage}[t]{0.9\columnwidth}
\begin{lstlisting}[language=python,alsoletter={.},deletekeywords={.sum,in},morekeywords={@proc,DRAM, f32,Neon,neon_vld_4xf32,neon_vst_4xf32,stage_mem, expand_dim,lift_alloc,autofission, replace, set_memory,bind_expr,reorder_loops,neon_vfmla_4xf32_4xf32, unroll_loop}]
# USER CODE
p = unroll_loop(p,'it')
p = unroll_loop(p,'jt')

# RESULTING EXO GENERATED CODE
def uk_8x12( KC: size, alpha: f32[1] @ DRAM, 
  Ac: f32[KC, 8] @ DRAM, Bc: f32[KC, 12] @ DRAM, 
  beta: f32[1] @ DRAM, C: f32[12, 8] @ DRAM
  ):
  
  # Registers for C and load C as in 
  # previous figures.  Omitted for brevity

  # Registers for Ac and Bc
  A_reg: R[2, 4] @ Neon
  B_reg: R[3, 4] @ Neon
  
  # C += Ac * Bc 
  for k in seq(0, KC):
    # Unrolled loads from Ac and Bc to registers
    neon_vld_4xf32(A_reg[0, 0:4], Ac[k, 0:4 + 0])
    neon_vld_4xf32(A_reg[1, 0:4], Ac[k, 4:4 + 4])
    neon_vld_4xf32(B_reg[0, 0:4], Bc[k, 0:4 + 0])
    neon_vld_4xf32(B_reg[1, 0:4], Bc[k, 4:4 + 4])
    neon_vld_4xf32(B_reg[2, 0:4], Bc[k, 8:4 + 8])
    # Computation with registers
    for jt in seq(0, 3):
      for it in seq(0, 2):
        for jtt in seq(0, 4):
          neon_vfmla_4xf32_4xf32(
            C_reg[jtt + 4 * jt, it, 0:4],
            A_reg[it, 0:4], B_reg[0:4, jt], jtt)
  
  # Store C from registers as in 
  # previous figures.  Omitted for brevity 
\end{lstlisting}
\end{minipage}
\caption{\exo code for \gemm micro-kernel v6.}
\label{lst:v6} 
\end{figure}

\paragraph{\textbf{Generated C-code}} To ensure that the generated code is not only optimized in the C language but also that the compilation to assembly is done correctly, we have compiled the c-code with the \texttt{gcc-10 -S} command, and the resulting assembly code for the $k$-loop is shown in Figure~\ref{lst:ass}. This output is as optimized as the one implemented by hand in BLIS.

\begin{figure}[tbh!]
\centering
\begin{minipage}[t]{0.9\columnwidth}
\begin{lstlisting}[language=c,alsoletter={.},deletekeywords={.sum,in},morekeywords={fmla, ldp, addr, ldr, cmp, bne, add}]
.L3:
  ldp     q1, q0, [x3]           #load A -> q0, q1
  add     x0, x0, 1       
  ldp     q4, q3, [x4]           #load B -> q3, q4
  add     x3, x3, 32      
  ldr     q2, [x4, 32]           #load B -> q2
  add     x4, x4, 48       
  fmla    v12.4s, v1.4s, v4.s[0] # C = C + q1 * q4[0]
  fmla    v10.4s, v1.4s, v4.s[1] # C = C + q1 * q4[1]
  fmla    v8.4s, v1.4s, v4.s[2]  # C = C + q1 * q4[2]
  fmla    v30.4s, v1.4s, v4.s[3] # ...
  fmla    v11.4s, v0.4s, v4.s[0] 
  fmla    v9.4s, v0.4s, v4.s[1]  
  fmla    v31.4s, v0.4s, v4.s[2] 
  fmla    v29.4s, v0.4s, v4.s[3] 
  fmla    v28.4s, v1.4s, v3.s[0] 
  fmla    v26.4s, v1.4s, v3.s[1] 
  fmla    v24.4s, v1.4s, v3.s[2] 
  fmla    v22.4s, v1.4s, v3.s[3] 
  fmla    v27.4s, v0.4s, v3.s[0] 
  fmla    v25.4s, v0.4s, v3.s[1] 
  fmla    v23.4s, v0.4s, v3.s[2] 
  fmla    v21.4s, v0.4s, v3.s[3] 
  fmla    v20.4s, v1.4s, v2.s[0] 
  fmla    v18.4s, v1.4s, v2.s[1] 
  fmla    v16.4s, v1.4s, v2.s[2]  
  fmla    v6.4s, v1.4s, v2.s[3]   
  fmla    v19.4s, v0.4s, v2.s[0]  
  fmla    v17.4s, v0.4s, v2.s[1]  
  fmla    v7.4s, v0.4s, v2.s[2]   
  fmla    v5.4s, v0.4s, v2.s[3]   
  cmp     x1, x0
  bne     .L3             
\end{lstlisting}
\end{minipage}
\caption{Assembly generated with the gcc-10 compiler of the \exo code for \gemm micro-kernel.}
\label{lst:ass} 
\end{figure}

\subsection{Edge Cases}

One of the problems with the one micro-kernel per architecture approach adopted by the HPC libraries is the performance degradation when the dimensions of the micro-kernel do not match the optimized ones. This situation is called an edge case. Solutions such as BLIS use a non-specialized version of the micro-kernel for these situations because the edge cases have no performance impact for large  problem sizes. However, newer HPC scenarios such as DL are full of these edge cases. A clear example is the sizes of the first layer of the ResNet50-v1.5 convolutional model, where after applying the \imr method, the resulting \gemm dimensions are  $12544, 64,$ and $147$ for $M, N,$ and $K$, respectively.

Using \exo, and assuming that the  BLIS packings are in the \gemm algorithm, all we need to do is use the code shown in Figure~\ref{lst:v1} and change the values for $M_R$ and $N_R$ to match the edge cases. Then, running all the steps will generate a new micro-kernel that matches those values. 

However, it is possible that we do not need the packing because the data is already packed or the size of the problem is small enough that the cost of packing is not worth it. In this scenario, we should adjust the procedure of the micro-kernel generation (e.g. non-packing of A) as follows:
\begin{enumerate}
    \item Loop \texttt{i} in Figure~\ref{lst:v2} should not be split.
    \item The mapping between $A_c$ and \texttt{A\_reg} will change and the dimensions of the latter will match the value of $M_R$.
    \item Inside the \texttt{k}-loop, \texttt{A\_reg} will be broadcasted with the values $A_c$.
    \item The calculation will use the \texttt{neon\_vfmadd\_4xf32\_4xf32}, which computes the broadcasted \texttt{A\_reg} by the entire \texttt{B\_reg}.
\end{enumerate}


\subsection{Architectural Portability}

\exo only needs to change the third argument in the \texttt{replace} statements in the user code to generate the desired code. If the new architecture provides an API with the same functionality, this is the required change.

However, it is possible that a statement used in this example is not present in the intrinsic Application Programming Interface
 (API) of other hardware (e.g., ARM Neon \texttt{vfmaq\_laneq\_f32}). Then, a similar approach to the one presented when the non-packing of data is available is used.

As a simple example, changing the line 21 in Figure~\ref{lst:v3} from \texttt{replace(p, ’for itt in \_: \_’, neon\_vld\_4xf32)} by \texttt{replace(p, ’for itt in \_: \_’, \_mm512\_loadu\_ps)} will change the load intrinsic from ARM Neon to Intel AVX512.

\subsection{Data types}
Generating micro-kernels for different data types is as easy as using the function \texttt{set\_precision} function for each memory allocation and register in the code. For example, \texttt{set\_precision(p, A\_reg,"f16")} will use 16~bit floating point registers for the \texttt{A\_reg} variable. Also, the \texttt{Neon} argument in the \texttt{set\_memory()} statements must be changed to \texttt{Neon8f}.

\section{Performance evaluation}
\label{sec:experiments}

In this section 
we evaluate the performance of the \gemm BLIS-like routines and $8 \times 12$ microkernels comparing three different microkernel implementations: \texttt{Neon}, a neon-intrinsic, hand-developed microkernel; \texttt{BLIS}, the BLIS~v.0.9 microkernel; and \exo, the automatically generated code presented in Section~\ref{sec:generator}. The experiments in this section were performed on a single core of the NVIDIA Carmel processor (ARM v8.2) embedded on an NVIDIA Jetson AGX Xavier board, using IEEE 32-bit floating point arithmetic (FP32).

The experiments include three types of \gemm problems:
microkernel performance in a solo mode (including the edge cases); large square matrices; and highly ``rectangular'' problems. 

\subsection{Solo Mode}

\begin{figure}[tb!]
\centering
\includegraphics[width=\columnwidth]{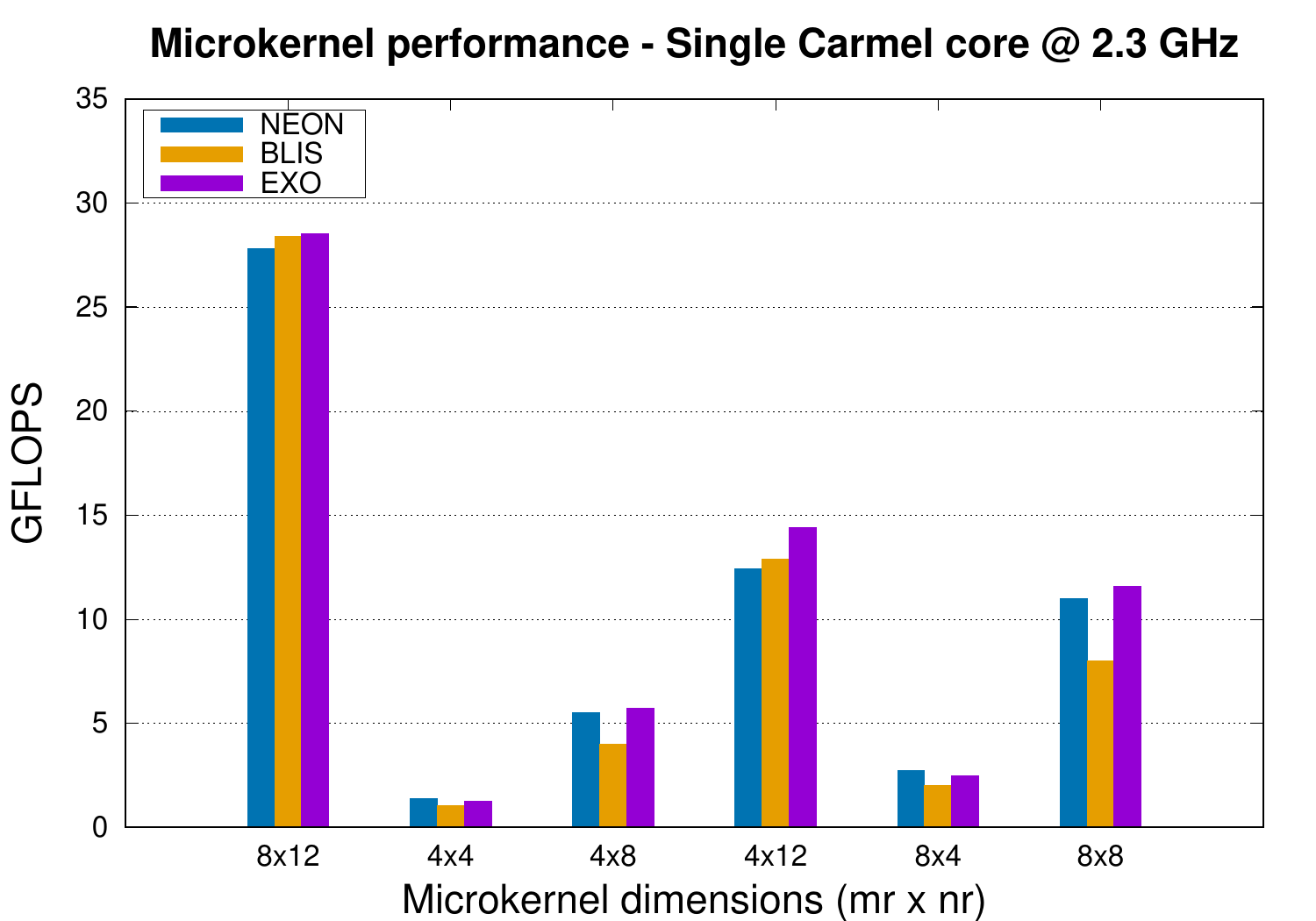}
\caption{Performance of different  micro-kernels in solo-mode with different problem sizes. \texttt{NEON}, and \texttt{BLIS} use the same micro-kernel for all scenarios while \texttt{EXO} uses an ad-hoc auto-generated micro-kernel for each one.}
\label{fig:solo}
\end{figure}

In this experiment, we show that the  \exo generated micro-kernel is as good as the hand-coded ones. To do this, we run the micro-kernels directly for 5~seconds and then compute the GFLOPS. The $8 \times 12$ columns in Figure~\ref{fig:solo} show the performance when the micro-kernel is invoked at its maximum performance. For this test, we have set the $K_c$ to 512, which is the value of BLIS packing for this ARM architecture. There are minor differences between the three solutions. First of all, \texttt{NEON} is slower than  \texttt{BLIS}, and the main difference is that the former is written with Neon intrinsics while the latter is in assembly. \exo is slightly better because it only supports the exact case of $8 \times 12$, while the other two micro-kernels also include the logic for the edge cases. 

The other columns represent the performance when calling the micro-kernels with different edge cases. While \texttt{NEON} and \texttt{BLIS} run the same micro-kernel that in the case of $8 \times 12$, \exo benefits from the easy way to generate different micro-kernel sizes and therefore a specialized micro-kernel is run for each size of the problem. This approach is clearly the best solution for overcoming edge cases.

\subsection{Squared Matrices}

\begin{figure}[tb!]
\centering
\includegraphics[width=\columnwidth]{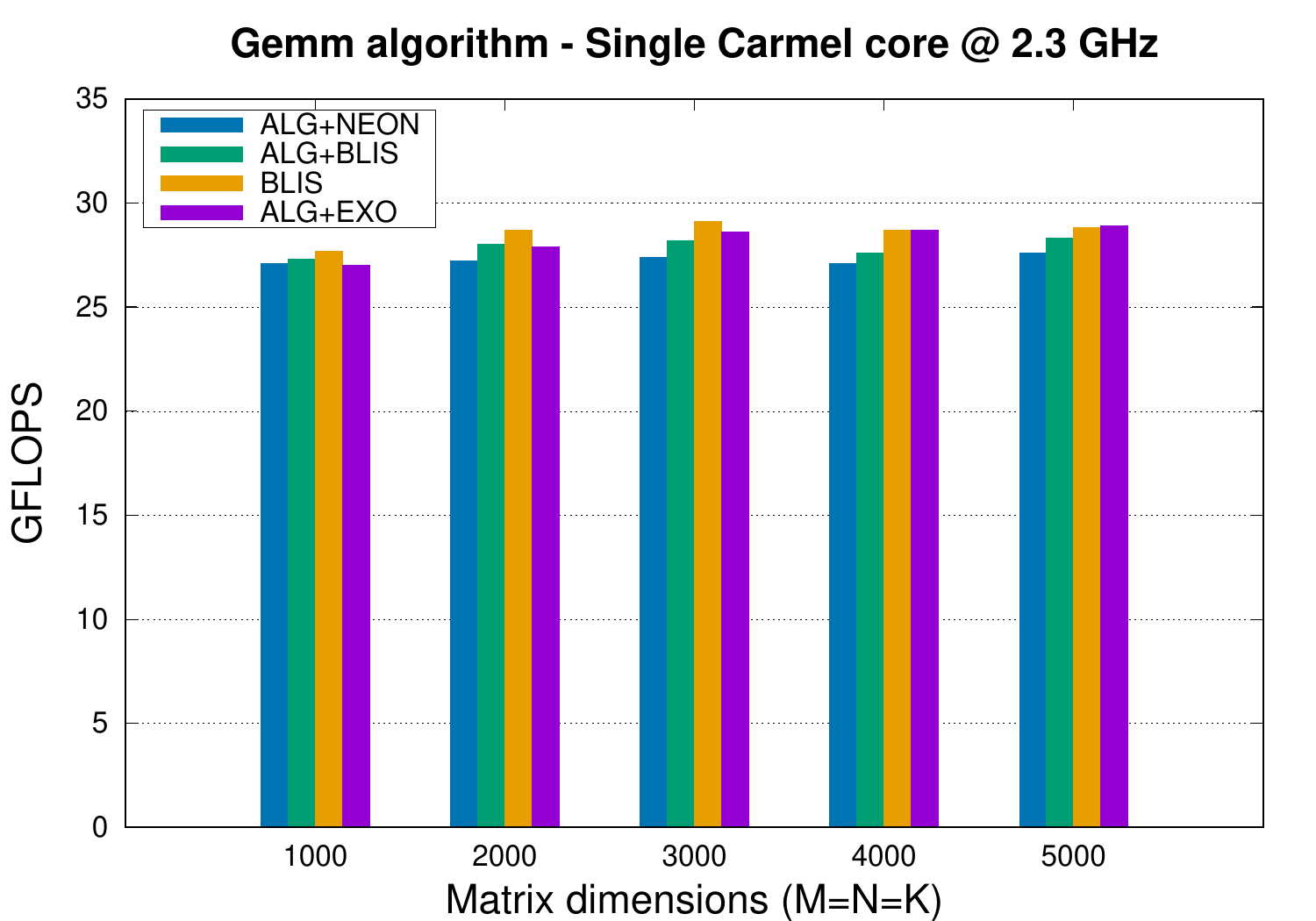}
\caption{Performance of different  micro-kernels for squarish \gemm. \texttt{NEON}, and \texttt{BLIS} use their unique micro-kernel while  \texttt{EXO} uses a set of auto-generated micro-kernels.}\label{fig:Square}
\end{figure}

Figure~\ref{fig:Square} shows the performance for a complete execution of the \gemm algorithm with the micro-kernels. The columns with the prefix \texttt{ALG+} indicate that we have used a BLIS-like realization of the \gemm algorithm, which also includes the theoretical model presented in~\cite{BLIS4} for optimizing the packings. The suffix of these columns indicates the employed micro-kernel (or micro-kernels for \exo) used. In addition, the column labeled as \texttt{BLIS} is the performance when calling the \gemm function of the BLIS library.

\texttt{BLIS} performs better in this case because the \gemm algorithm used in the BLIS library implements prefetching inside the micro-kernel that is not used in the \texttt{ALG+BLIS} approach. \texttt{ALG+\exo} outperforms other \texttt{ALG+} and considering that the packings of the micro-kernel operands are equal due to the model, the only difference is the use of different micro-kernels (\exo) or one micro-kernel for all cases. Specifically, for the \exo approach we have used the $8 \times 4$ micro-kernel for the 1,000 and 4,000 problem sizes and the $8 \times 8$ micro-kernel for the 2,000 and 5,000 problem sizes.

\subsection{Rectangular Matrices}
Given the current interest in DL inference, the dimensions of this experiment are chosen to be those obtained by applying the 
\imr transform~\cite{Che06} to the convolution layers in the ResNet50 v1.5 and VGG16 deep neural 
network (DNN) models in the form of a \gemm. The ``batch'' size for the inference scenario is set 
to 1 sample.
Since some layers share the same parameters, resulting in \gemm problems 
of the exact dimensions, 
we report the results for these only once; see Table~\ref{table:resnet50} and Table~\ref{table:vgg16} for reference.

\begin{table}[]
\vspace*{2ex}
\centering
\footnotesize

\begin{tabular}{rlrrr}
\toprule

Layer id. & Layer numbers ResNet50 v1.5& $m$ & $n$ & $k$  \\ 
  \cline{1-5}
~1&001 &                     12,544 &  ~~64 & ~147  \\
~2&006 &                     ~3,136 &  ~~64  & ~~64  \\
~3&009/021/031 &             ~3,136 &  ~~64  & ~576  \\
~4&012/014/024/034 &         ~3,136 &  ~256  & ~~64  \\
~5&018/028 &                 ~3,136 &  ~~64  & ~256 \\
~6&038 &                     ~3,136 &  ~128  & ~256 \\
~7&041/053/063/073 &         ~~784 &  ~128  & 1,152 \\
~8&044/056/066/076 &         ~~784 &  ~512  & ~128 \\
~9&046 &                     ~~784 &  ~512  & ~256 \\
10&050/060/070 &             ~~784 &  ~128  & ~512 \\
11&080 &                     ~~784 &  ~256  & ~512 \\
12&083/095/105/115/125/135 & ~~196 &  ~256  & 2,304 \\
13&086/098/108/118/128/138 & ~~196 &  1,024  & ~256 \\
14&088 &                     ~~196 &  1,024  & ~512 \\
15&092/102/112/122/132 &     ~~196 &  ~256  & 1,024 \\
16&142 &                     ~~196 &  ~512  & 1,024\\
17&145/157/167 &             ~~~49 &  ~512  & 4,608 \\
18&148/160/170 &             ~~~49 &  2,048  & ~512 \\
19&150 &                     ~~~49 &  2,048  & 1,024 \\
20&154/164 &                 ~~~49 &  ~512  & 2,048 \\ \bottomrule
\end{tabular}
\caption{Dimensions of the \gemm resulting from applying the \imr transform to the layers of the ResNet50 v1.5 DNN model with a batch size of 1 sample.}
\label{table:resnet50}
\end{table}

\begin{table}[]
\vspace*{2ex}
\centering
\footnotesize
\begin{tabular}{rlrrr}
\toprule

Layer id. & Layer numbers VGG16 & $m$ & $n$ & $k$  \\ 
  \cline{1-5}
~1&01 &                     50,176 &  ~64  & ~~27  \\
~2&03 &                     50,176 &  ~64  & ~576  \\
~3&06 &                     12,544 &  128  & ~576  \\
~4&08 &                     12,544 &  128  & 1,152  \\
~5&11 &                 ~3,136 &  256  & 1,152 \\
~6&13/15 &                     ~3,136 &  256  & 2,304 \\
~7&18 &         ~~784 &  256  & 2,304 \\
~8&20/22 &         ~~784 &  512  & 4,608 \\
~9&25/27/29&                     ~~196 &  512  & 4,608 \\ \bottomrule
\end{tabular}
\caption{Dimensions of the \gemm resulting from applying the \imr transform to the layers of the VGG16 DNN model with a batch size of 1 sample.}
\label{table:vgg16}
\end{table}

\begin{figure}[tb!]
\centering
\includegraphics[width=\columnwidth]{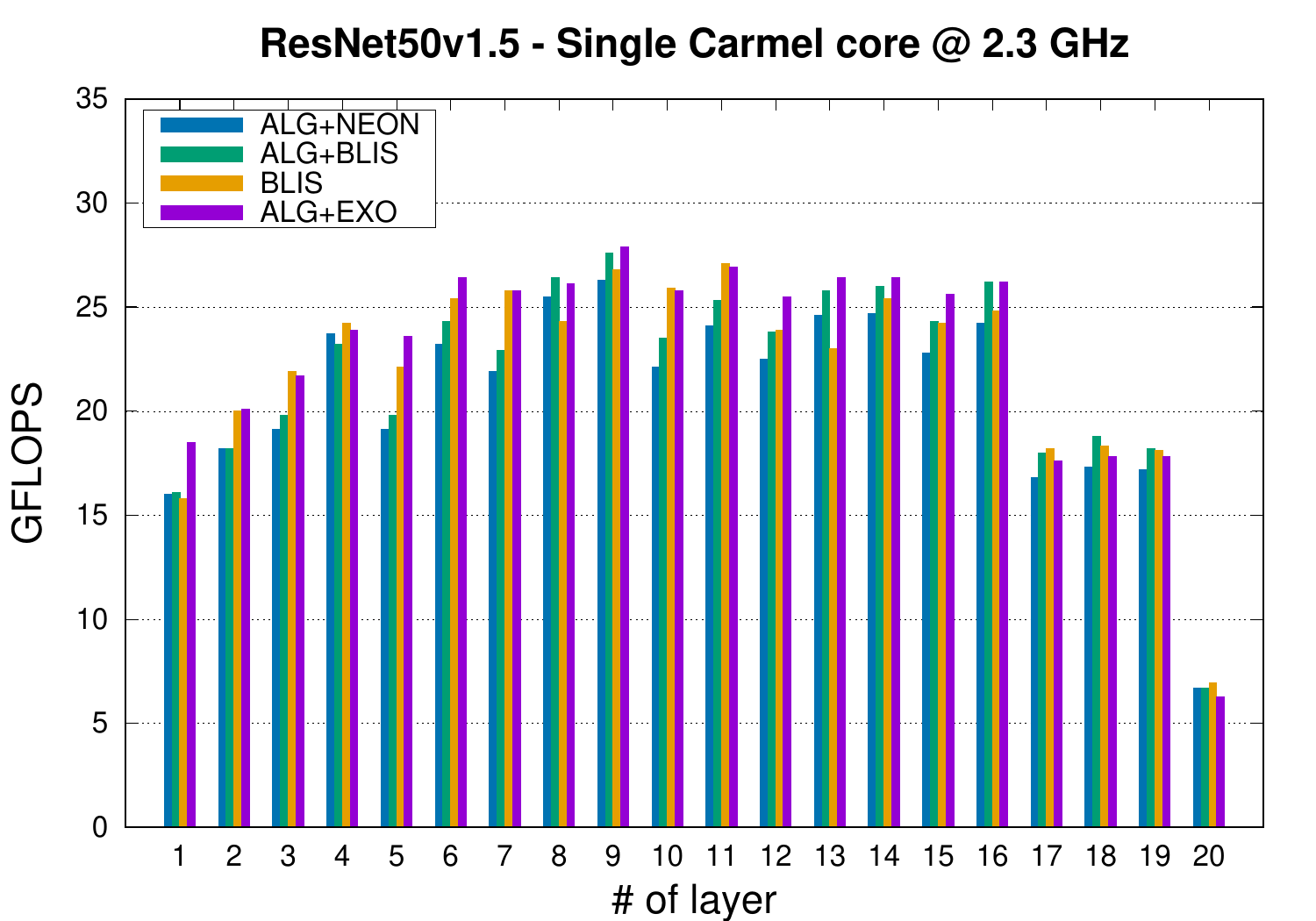}
\caption{Performance per layer of  ResNet50 v1.5 model.}
\label{fig:resnet}
\end{figure}

Figure~\ref{fig:resnet}  reflects the advantages of ad-hoc micro-kernels for edge cases. The \texttt{ALG+\exo} implementation is the best option for 9 out of 20 layers for ResNet50 v1.5, while \texttt{BLIS} with prefetching is the best for 6 of them. For this execution,  \texttt{ALG+\exo} uses the micro-kernels $8 \times 12$, $8 \times 4$, $4 \times 4$, $4 \times 8$, $4 \times 12$, $1 \times 8$, and $1 \times 12$. 

\begin{figure}[tb!]
\centering
\includegraphics[width=\columnwidth]{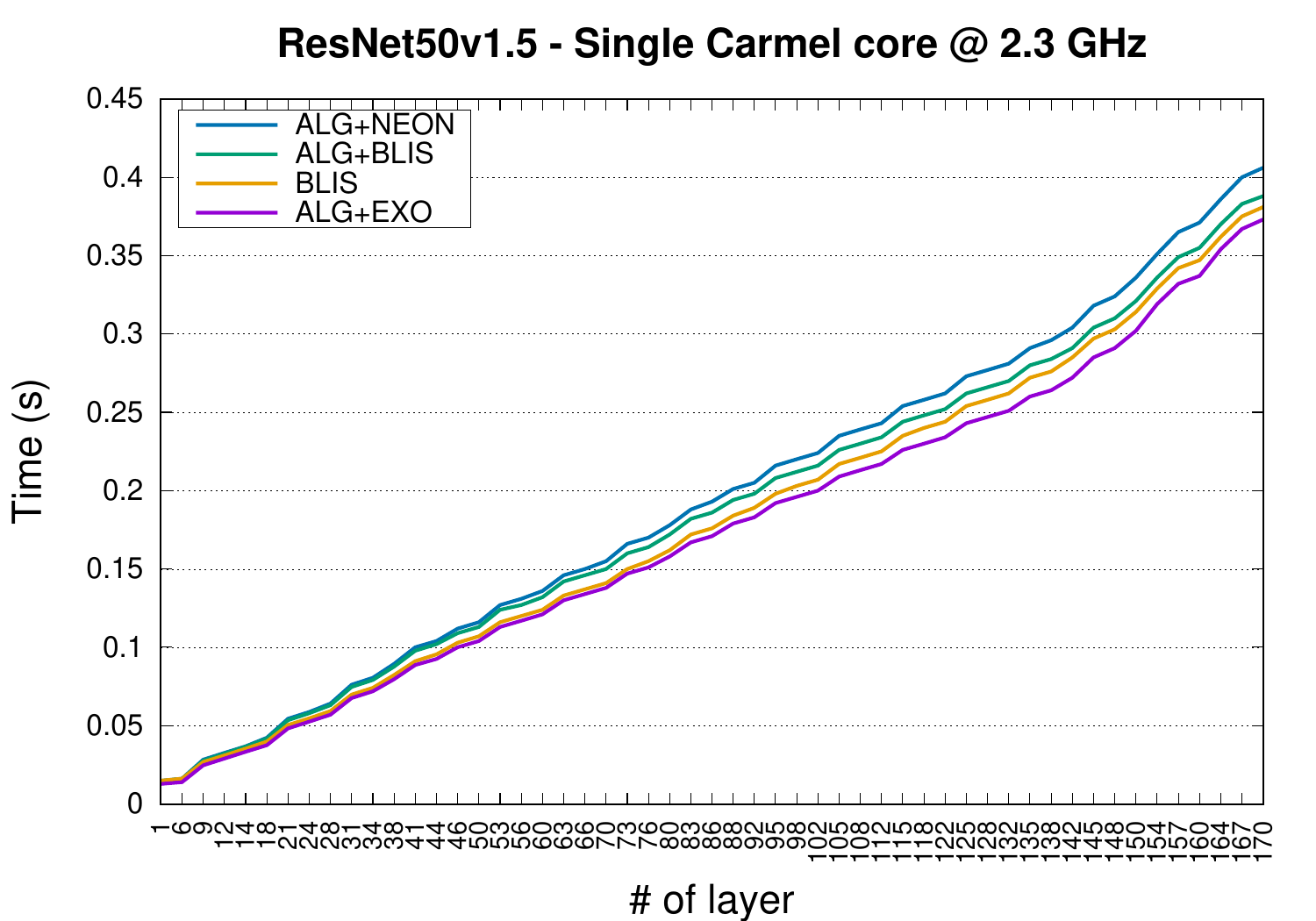}
\caption{Aggregated inference time for ResNet50 v1.5 model.}
\label{fig:resnet_agg}
\end{figure}

\begin{figure}[tb!]
\centering
\includegraphics[width= \columnwidth]{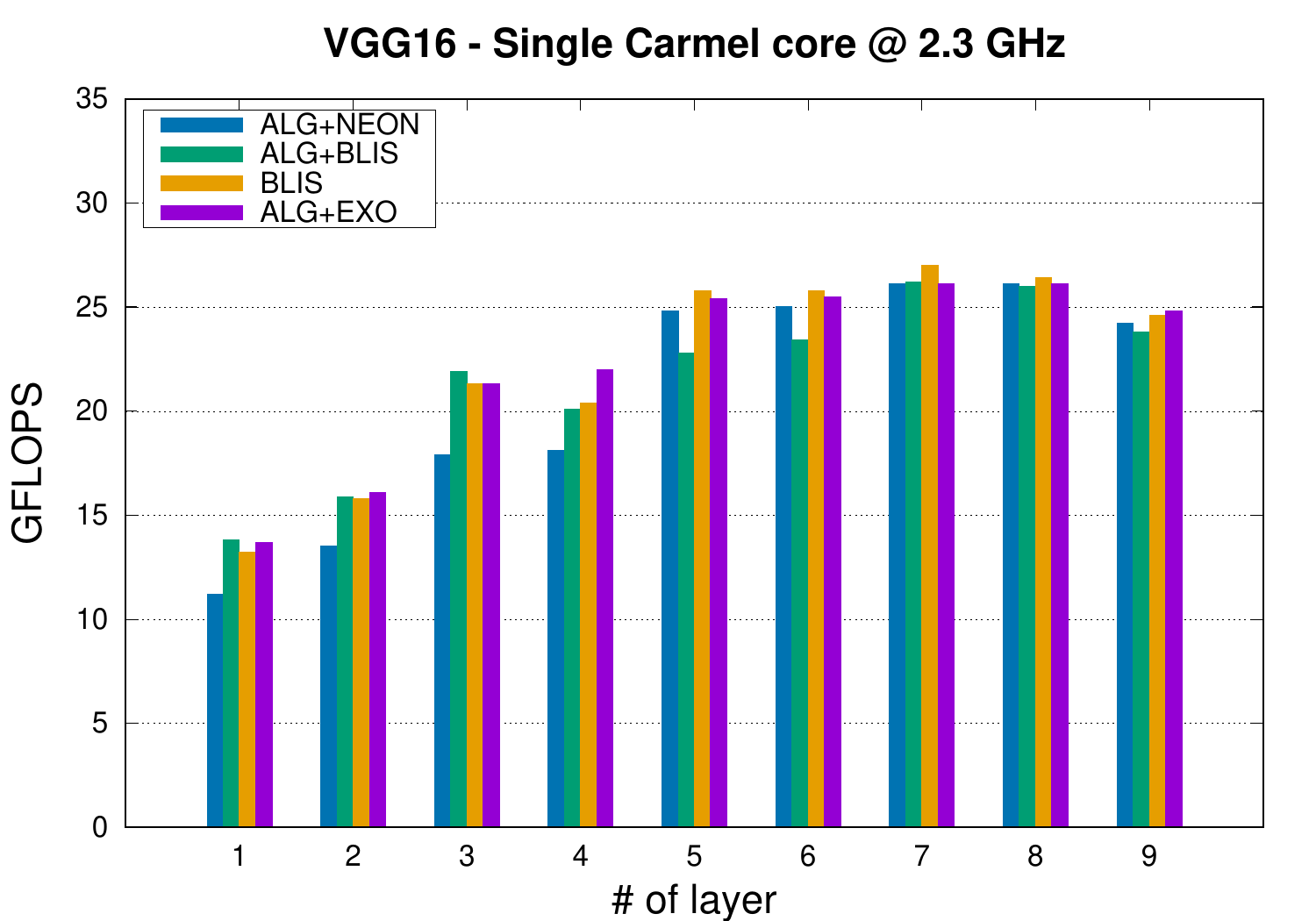}
\caption{Performance per layer of  VGG16 model.}
\label{fig:vgg}
\end{figure}

\begin{figure}[tb!]
\centering
\includegraphics[width=\columnwidth]{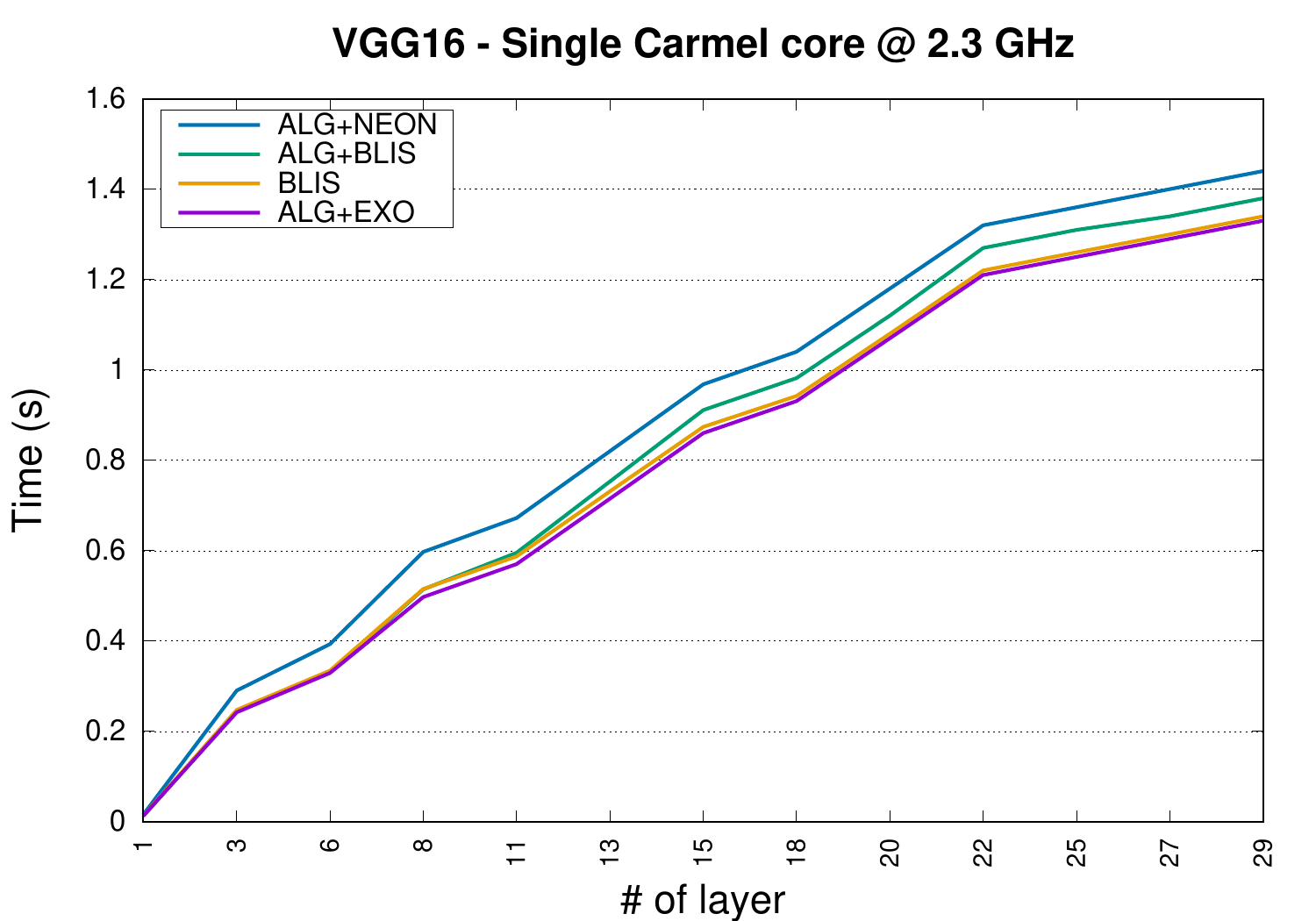}
\caption{Aggregated inference time for  VGG16 model.}
\label{fig:vgg_agg}
\end{figure}

To put these results in terms of absolute performance, Figure~\ref{fig:resnet_agg} shows the aggregated time for the entire model execution. Although the difference is small the best performance is achieved by \texttt{ALG+\exo}, followed by \texttt{BLIS}, \texttt{ALG+BLIS}, and \texttt{ALG+Neon}. Figure~\ref{fig:vgg} also improves the use of \exo, which is the best for 3 layers,  \texttt{BLIS} with prefetching in the other four of them, while the \texttt{ALG+BLIS} is the best on two of them. In terms of aggregated time, Figure~\ref{fig:vgg_agg} shows that the performance of \texttt{ALG+\exo} and \texttt{BLIS} solutions are close. 

\section{Conclusions}
\label{sec:conclusions}

In this paper, we have addressed the problem of the monolithic approach taken by LA libraries by generating specific, hardware-oriented micro-kernel C-code using the \exo tool. 
We have described in a step-by-step process how to build a micro-kernel generator from scratch that produces a C-code that performs close to (or even better than) the well-known code in the BLIS library. We have also provided hints for adapting the micro-kernel generator to meet  different software requirements such as different data types. We have analyzed this performance comparison in three different scenarios: micro-kernel execution, large, squarish matrix multiplications, and rectangular \gemm generated by the DL convolutional models against Neon intrinsics-based and assembly micro-kernels. 
In addition, this work has contributed to the \exo tool code with the support for some ARM features. 

As future work, {\color{black} we will adapt and analyze the micro-kernel generator tool with other architectures such as Intel, RISC-V, or the matrix engine}. In addition, we plan to tackle the generation of other pieces of code  for LA libraries or specific domain codes such as convolutional codes. 

\section*{Acknowledgments}
We thank Prof. Gilbert Bernstein from the University of Washington and Prof. Jonathan Ragan-Kelley from Massachusetts Institute of Technology for their active collaboration. A. Castell\'o is a FJC2019-039222-I fellow
supported by MCIN/AEI/10.13039/501100011033. Y. Ikarashi is supported by the Funai Overseas Scholarship, Masason Foundation, and Great Educators fellowships.
H. Mart\'inez is a postdoctoral fellow supported by the \emph{Consejer\'ia de Transformaci\'on Econ\'omica, Industria, Conocimiento y Universidades de la Junta de Andaluc\'ia}.

\newpage

%
%
%
%
%


\appendix
\section{Artifact Appendix}

\subsection{Abstract}

This appendix documents the EXO\_ukr\_generator software artifact and the procedure of how to install and execute it for the reproduction of the results shown in the paper. This software package is aimed at reproducing the overall sections of the paper including both the step-by-step micro-kernel design and the experimental results. This software is designed and configured for ARMv8 processors because it is the hardware used in the paper. However, it can be executed over ARM processors with the support of Neon intrinsics instructions The artifact is publicly available at {\url{https://github.com/adcastel/CGO_paper44_artifact}} and includes a series of scripts for building the environment and executing the experimentation.

\subsection{Artifact check-list (meta-information)}

{\small
\begin{itemize}
  \item {\bf Algorithm: }High-performance matrix multiplication
  \item {\bf Compilation: }Exo compiler
  \item {\bf Hardware: } ARM Neon (v8)
  \item {\bf Metrics:  } GFLOPS
  \item {\bf Experiments: } Stand-alone micro-kernels and deep neural networks \gemm
  \item {\bf How much time is needed to prepare workflow (approximately)?: }20 minutes
  \item {\bf How much time is needed to complete experiments (approximately)?: } 10 minutes
  \item {\bf Publicly available?: Available at } {\url{https://github.com/adcastel/CGO_paper44_artifact}}
  \item {\bf Code licenses (if publicly available)?: } None
\end{itemize}

\subsection{Description}

\subsubsection{How delivered}
This artifact is available via the {\url{https://github.com/adcastel/CGO_paper44_artifact}} Github repository without any license. This feature is indicated in the REQUIREMENTS.txt file.
\\

\subsubsection{Hardware dependencies}
The artifact is available for ARM v8 (Neon instructions support) and it has been tested in NVIDIA Xavier, Orin, and Nano platforms. This feature is indicated in the REQUIREMENTS.txt file.
\\

\subsubsection{Software dependencies}
This artifact comprises Python and C code, and therefore, versions of \texttt{Python3.9} and \texttt{gcc-10} (or higher) are mandatory. It is possible that there are some Python packages that are not currently installed. This possible scenario has been treated in the configuration scripts, however, package failure may occur. 
The software for the Exo and the Blis libraries is also included in the package.
This feature is indicated in the REQUIREMENTS.txt file.
For the plotting procedure, this artifact uses \texttt{gnuplot} tool.
\\

\subsubsection{Data sets}
The data sets for the experimentation are included in the repository code so there is not any extra requirement for them.
\subsection{Installation}
\begin{enumerate}
    \item 
For the installation, we first need to clone the repository with the command:
\\

\texttt{git clone \url{https://github.com/adcastel/CGO_paper44_artifact.git}}
\\
    \item 

Then we need to enter the directory with:
\\

\texttt{cd CGO\_paper44\_artifact}
\\
    \item 

and execute the \texttt{build.sh} script as:
\\

\texttt{source build.sh}
\\

This script will check the existent compilers and will build and install the Blis and the Exo software as well as set the environment variables.
\end{enumerate}

\subsection{Experiment workflow}

For the micro-kernel generation, we only need to execute the script as follows:
\\

\texttt{./microkernel\_generator.sh}
\\

This script will generate the micro-kernel explained in Section III of the paper, showing the generated code in each step of the process as shown in the section mentioned above. 

\subsection{Evaluation and expected result}

For the evaluation reproduction, we need to use two different scripts corresponding to the different evaluations in the paper.
\\

\begin{enumerate}

\item
For the evaluation of the micro-kernel in solo-mode we need to execute the corresponding script as follows:
\\

\texttt{./execute\_ukernel\_solo.sh}
\\

This script will execute the experiments shown in Figure~\ref{fig:solo}.
\\

\item

For reproducing the experiments of figures~\ref{fig:Square}-\ref{fig:vgg_agg} we should execute the following script as:
\\

\texttt{./execute\_algorithm.sh}
\\

This evaluation is the one that consumes more time due to the different combinations of algorithm, micro-kernel, and \gemm sizes (including both square and DL models). 
\\

\item 

For generating the plots, we should execute the plotting script as follows:
\\

\texttt{./do\_plots.sh}
\\

The plots of the paper will be located in the \texttt{plots} folder.
\\

\end{enumerate}

Please, notice that the user can build and execute the artifact using just one script:
\\

\texttt{./build\_and\_execute\_all.sh}
\\

This script will build the environment and execute all the experiments on their own.
\subsection{Experiment customization}

This artifact can be customized in several ways. The user is able to modify the generated micro-kernel or the evaluation of the \gemm.
\\

In the case of the micro-kernel generation, the user should modify the \texttt{generator.py} file inside the \texttt{EXO\_ukr\_generator} directory. There, the user can set the values of $M_R$ and $N_R$ (that are the sizes of the micro-kernel) and additionally the datatype. Please, notice that in that file, the first call is the one used in Section III of the paper while there are other commented lines that can be used for generating different micro-kernel sizes.
\\

If the user wants to change the experimental setup for the algorithm+micro-kernel evaluation, one needs to change the files in \texttt{CGO\_paper44\_artifact/gemm\_blis\_family
/cnn\_models/} or to generate a new input file and then add the corresponding line to the \texttt{execute\_algorithm.sh} file.
\\

Please, notice that the overall evaluation of a different micro-kernel (of those that are used in the paper) will involve different changes in every test\slash configuration. Therefore, the ``random'' attempts with non-usual micro-kernel sizes are not recommended without the guidance of the artifact developers.



%
\IEEEpeerreviewmaketitle



\ifCLASSOPTIONcaptionsoff
  \newpage
\fi


\bibliographystyle{IEEEtran}
\bibliography{paper}

\vfill


\end{document}